\documentclass[10pt,conference]{IEEEtran}

\IEEEoverridecommandlockouts

\usepackage{cite}
\usepackage{amsmath,amssymb,amsfonts}
\usepackage{algorithmic}
\usepackage{graphicx}
\usepackage{textcomp}
\usepackage{xcolor}
\usepackage{booktabs}
\usepackage{multirow}
\usepackage{hyperref}
\usepackage{balance}
\usepackage{tikz}
\usetikzlibrary{positioning, arrows.meta, calc}

\begin{document}

\title{Family-Aware Residual Architecture for Predicting\\Quantum Circuit Simulation Performance\thanks{\textcopyright~2026 IEEE. Personal use of this material is permitted. Permission from IEEE must be obtained for all other uses, in any current or future media, including reprinting/republishing this material for advertising or promotional purposes, creating new collective works, for resale or redistribution to servers or lists, or reuse of any copyrighted component of this work in other works. Accepted as a full paper at IEEE ISVLSI 2026 (QC-CSAA Workshop). To appear in IEEE Xplore.}}

\author{
\IEEEauthorblockN{Honjar Xing, Yehong Jiang, Xianbang Wang, Zehua Wang, Zhicheng Jiang}
\IEEEauthorblockA{Massachusetts Institute of Technology\\
Cambridge, MA 02139, USA\\
\{honjar, yehong28, xbwang, zehuaw, jzc\_2007\}@mit.edu}
}

\maketitle

\begin{abstract}
Approximate tensor-network simulators enable classical simulation of quantum circuits beyond the reach of exact methods, but selecting optimal approximation parameters---such as bond dimension thresholds---remains a costly trial-and-error process. We present a family-aware neural architecture that predicts both the minimum approximation threshold required to achieve target fidelity and the expected wall-clock runtime for quantum circuit simulation, given only the circuit's OpenQASM description and execution context. Our key insight is that quantum circuits from different algorithmic families (e.g., QFT, Grover, VQE) exhibit fundamentally distinct simulation cost profiles due to their differing entanglement structures. We employ family-conditioned residual corrections---additive, family-specific adjustments atop a shared backbone, drawing on established conditional computation techniques---enabling the model to capture both universal circuit properties and algorithmic nuances. The architecture incorporates a pretrained family classifier (97.5\% accuracy) and domain-informed algorithm fingerprint features derived from gate-composition heuristics. Evaluated on circuits spanning 7--130 qubits across 10 algorithm families, our system achieves 79.5\% exact threshold accuracy (91.2\% within one rung) and $R^2=0.82$ runtime correlation, with inference completing in approximately 50\,ms---replacing trial-and-error simulation runs that may take minutes to hours. Ablation studies confirm that family-aware modeling provides the single largest performance improvement (+3.2 percentage points), validating the hypothesis that algorithm family is a first-class feature for simulation cost prediction.
\end{abstract}

\section{Introduction}
\label{sec:introduction}

Quantum circuit simulation is indispensable for algorithm development, verification, and benchmarking, yet exact statevector simulation scales exponentially with qubit count, rendering circuits beyond roughly 40--50 qubits intractable on classical hardware~\cite{markov2008,nielsen2010}. Approximate methods based on tensor network contraction~\cite{markov2008,pan2024,huang2021,nguyen2022} extend the frontier substantially by truncating bond dimensions to bound computational cost, but they introduce a parameter selection problem: practitioners must choose an approximation threshold that balances fidelity against runtime. In practice, this selection is trial-and-error---a developer guesses a threshold, runs a simulation that may take minutes to hours, discovers that the result is either insufficiently accurate or wastefully conservative, and iterates. In our benchmarks (Section~\ref{sec:experiments}), individual simulation runs ranged from under one second for small circuits at low bond dimensions to over an hour for large circuits, and finding suitable parameters typically requires several such attempts. Bad choices squander significant computational resources or produce unusable outputs.

The core challenge we address is the following: given a quantum circuit specified in OpenQASM and an execution context (hardware backend and numerical precision), predict (1)~the minimum approximation threshold required to achieve a target fidelity and (2)~the expected wall-clock runtime, \emph{without actually running the simulator}. This is fundamentally a resource estimation problem for approximate quantum simulation. An accurate predictor would allow developers to select parameters confidently before committing compute, transforming an expensive search process into a single forward pass of a lightweight model.

Our key insight is that circuits from different quantum algorithm families exhibit fundamentally different simulation cost profiles. QFT circuits possess structured entanglement patterns that tensor networks can exploit efficiently. Grover's search circuits feature amplitude amplification with characteristic Hadamard--oracle--diffusion structures. Random circuits, by contrast, produce near-maximal entanglement that resists low-rank approximation. A single model that ignores these family-specific patterns leaves significant predictive signal on the table.

We make three contributions. First, we demonstrate that \emph{algorithm family is a first-class predictive feature} for simulation cost and validate this hypothesis through a family-aware architecture that applies conditional residual corrections~\cite{he2016} and Feature-wise Linear Modulation (FiLM)~\cite{perez2018film} atop a shared backbone, yielding the single largest accuracy improvement in our ablation study (+3.2 percentage points). Second, we introduce \emph{algorithm fingerprint features}---hand-crafted heuristic scores derived from gate-composition analysis that detect signatures of common quantum algorithms directly from the circuit description, designed for the small-data regime where learned feature extractors are infeasible. Third, we evaluate the approach on circuits spanning 7--130 qubits across 10 algorithm families using circuit-stratified cross-validation, and show that the system achieves 79.5\% exact threshold accuracy, 91.2\% within-one-rung accuracy, and $R^2=0.82$ runtime correlation at roughly 50\,ms inference latency on a single CPU core. Section~\ref{sec:related} reviews related work, Section~\ref{sec:method} details our method, Section~\ref{sec:experiments} presents experiments, and Section~\ref{sec:conclusion} concludes. This work directly addresses performance estimation and resource consumption prediction for approximate quantum simulation, core topics of the QC-CSAA workshop.

\section{Related Work}
\label{sec:related}

\subsection{Approximate Quantum Circuit Simulation}

Tensor network methods are the dominant paradigm for simulating quantum circuits beyond statevector limits. Markov and Shi~\cite{markov2008} established the theoretical foundations by framing circuit simulation as tensor network contraction, demonstrating that the computational cost is governed by the treewidth of the underlying graph. Subsequent work has focused on scaling these methods: Huang et~al.~\cite{huang2021} introduced efficient parallelization strategies for tensor network contraction that enabled simulation of circuits with hundreds of qubits, while Nguyen et~al.~\cite{nguyen2022} demonstrated exascale tensor network simulation through the TNQVM framework. More recently, Pan et~al.~\cite{pan2024} achieved substantial speedups by optimizing tensor network contraction for modern GPU architectures.

These advances improve the simulation engine itself but do not address the \emph{meta-problem} that motivates our work: predicting, before execution, which approximation settings will yield acceptable fidelity at reasonable cost. The parameter selection problem grows more acute as simulators become faster and are applied to larger circuits, because the space of possible configurations expands accordingly.

\subsection{Machine Learning for Quantum Computing Tooling}

Machine learning is increasingly applied to optimize quantum computing workflows. The most closely related prior work is that of Quetschlich et~al.~\cite{quetschlich2023pred}, who train ML models to predict which compilation options will produce the best results for a given quantum circuit; this line of work has since been extended to automatic device selection with device-specific compilation in the MQT Predictor system~\cite{quetschlich2025predictor}. Their approach treats circuits as fixed-length feature vectors extracted from structural properties and trains classifiers over compiler configurations. In a complementary line, Quetschlich et~al.~\cite{quetschlich2023dac} apply reinforcement learning to compiler optimization, learning sequential transformation policies. Wang et~al.~\cite{wang2022quest} propose a graph transformer for quantum circuit reliability estimation, demonstrating that graph-based representations can capture circuit structure for property prediction. The broader MQT ecosystem~\cite{wille2024} provides additional tooling for design automation in quantum computing.

Our work differs from these efforts in two respects. First, we target \emph{simulation} performance prediction rather than compilation quality---a distinct task with different output semantics (approximation thresholds and runtimes versus compiler flag selections). Second, we present an architecture that treats algorithm family as a first-class structural feature via conditional residual corrections and FiLM modulation~\cite{perez2018film}, drawing on established techniques from multi-task and conditional computation but applying them to a new domain where the conditioning signal---algorithm family---carries strong predictive power for simulation cost. A distinguishing challenge of our setting is data scarcity: while prior ML-for-quantum-computing work operates on hundreds to thousands of circuits~\cite{quetschlich2023pred,quetschlich2025predictor}, approximate simulation benchmarks are expensive to generate---each circuit--configuration pair requires a full simulation run whose cost ranges from seconds to hours depending on circuit size and approximation settings---making small-data effectiveness a design requirement rather than merely a limitation. Our family-aware conditioning and hand-crafted fingerprint features are explicitly motivated by this constraint.

\subsection{Quantum Circuit Benchmarking}

Standardized benchmark suites are essential for evaluating both quantum hardware and classical simulation tools. The MQT Bench suite~\cite{quetschlich2023bench} provides a systematic collection of quantum circuits organized by algorithm family, covering implementations of common algorithms (Grover, QFT, VQE, and others) across varying qubit counts. We leverage MQT Bench circuits as training data for our family classifier, exploiting its family labels as supervised signal.

Our work is situated in the NISQ era~\cite{preskill2018}, where noisy intermediate-scale devices coexist with classical simulators as primary tools for algorithm development. As circuits grow in size and diversity, the need for intelligent resource estimation---predicting simulation cost before committing to execution---becomes a practical bottleneck that our method directly addresses.

\section{Method}
\label{sec:method}

\subsection{Problem Formulation}
\label{sec:problem}

Given a quantum circuit $C$ described in OpenQASM and an execution context $e \in \{\text{CPU}, \text{GPU}\} \times \{\text{single}, \text{double}\}$ specifying backend and floating-point precision, we address two prediction tasks simultaneously.
First, we predict the minimum approximation threshold $\theta^{*} \in \{\star,\allowbreak 1, 2, 4, 8, 16, 32, 64, 128, 256\}$ required for an approximate tensor-network simulator~\cite{markov2008,kasirajan2024} to achieve fidelity $\geq 0.75$ (a standard target balancing simulation accuracy against computational cost), where each integer value controls the maximum bond dimension during tensor contraction (higher values permit more accurate but slower simulation), and $\star$ denotes that no threshold achieves the target.
Second, we predict the expected wall-clock runtime $r$ in seconds for a given circuit--threshold pair.

\subsection{Feature Extraction}
\label{sec:features}

We extract 32 numerical features from each circuit using Qiskit~\cite{qiskit2023}, organized into seven categories.
\emph{Basic Statistics} (3 features) capture circuit depth, qubit count, and total gate count.
\emph{Gate Type Counts} (12 features) record individual counts for each supported gate type (H, X, Y, Z, S, T, R$_x$, R$_y$, R$_z$, CX, CZ, SWAP).
\emph{Execution Context} (4 features) encodes backend and precision as binary indicators.
\emph{Complexity Metrics} (3 features) include the two-qubit gate ratio, qubit connectivity (graph density of the interaction graph), and the depth-to-width ratio.
\emph{Local Features} (2 features) capture the maximum degree and the Shannon entropy of the degree distribution in the qubit interaction graph.
\emph{Algorithm Fingerprints} (3 features), described in Section~\ref{sec:fingerprints}, provide heuristic scores detecting common algorithmic patterns.
\emph{Graph Optimizer Features} (5 features) summarize structural properties obtained after Reverse Cuthill--McKee reordering, including clustering coefficient, connected component count, maximum and mean cut counts across consecutive partitions, and maximum qubit-index span of two-qubit gates.

For unbounded features (e.g., gate counts, depth), we apply logarithmic scaling $\tilde{x} = \log(x + 1)$ to compress their dynamic range before model input. Bounded features such as ratios and densities are used directly.

\subsection{Algorithm Fingerprint Features}
\label{sec:fingerprints}

A key observation is that many quantum algorithms produce characteristic gate-composition signatures that are informative for simulation cost, yet invisible to generic circuit statistics.
We design three heuristic fingerprint scores---hand-crafted, not learned---that detect common algorithmic patterns directly from gate counts and circuit structure. We opt for hand-crafted heuristics over learned representations because the small training set (36 circuits) makes learning robust feature extractors infeasible, while domain-informed scores provide reliable signal.

\textbf{Arithmetic score.} Measures the density of multi-controlled gates (CCX, MCX, CSWAP) with a multiplicative boost when T/T$^{\dagger}$ gate density exceeds a threshold, capturing modular arithmetic subcircuits.

\textbf{QFT score.} Responds to controlled-phase gates (CP, CU1, MCP) combined with Hadamard gates, the defining signature of the Quantum Fourier Transform.
Penalties are applied when arithmetic or parameterized gate densities are high, separating pure QFT from QPE or Shor subcircuits.

\textbf{QNN score.} Measures parameterized-gate density (gates with non-trivial continuous parameters), suppressed when arithmetic structure dominates.
This score activates for variational circuits such as VQE ans\"{a}tze and quantum neural networks.

These fingerprints provide the model with coarse algorithmic identity signals without requiring explicit family labels, and they serve as informative features for both the family classifier and the performance predictor.

\subsection{Architecture}
\label{sec:architecture}

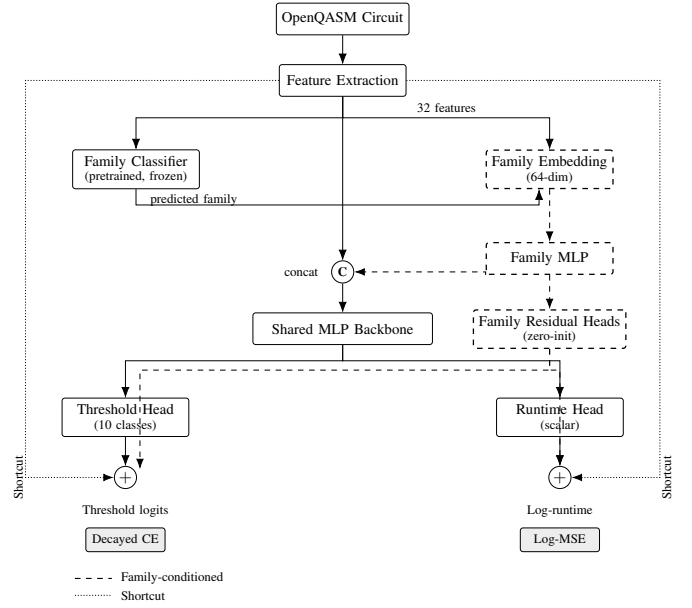
\begin{figure}[t]
\centering
\resizebox{\columnwidth}{!}{%
\begin{tikzpicture}[
    node distance=0.5cm,
    >={Latex[length=2mm]},
    box/.style={draw, rectangle, rounded corners=1.5pt,
        minimum height=0.6cm, minimum width=2.4cm,
        font=\footnotesize, align=center,
        inner sep=3pt, line width=0.4pt},
    novel/.style={draw, dashed, rectangle, rounded corners=1.5pt,
        minimum height=0.6cm, minimum width=2.4cm,
        font=\footnotesize, align=center,
        inner sep=3pt, line width=0.7pt},
    addnode/.style={draw, circle, minimum size=0.42cm,
        font=\small\bfseries, inner sep=0pt, line width=0.5pt},
    lossbox/.style={draw, rectangle, rounded corners=1.5pt,
        minimum height=0.45cm, minimum width=1.5cm,
        font=\scriptsize, align=center,
        inner sep=2pt, line width=0.4pt, fill=black!7},
    arr/.style={->, line width=0.4pt},
    darr/.style={->, dashed, line width=0.55pt},
    dotarr/.style={->, densely dotted, line width=0.55pt},
    lbl/.style={font=\scriptsize}
]

\node[box] (qasm) {OpenQASM Circuit};
\node[box, below=0.55cm of qasm] (feat) {Feature Extraction};
\draw[arr] (qasm) -- (feat);
\node[lbl, right=0.08cm of feat.south east, anchor=north west] {32 features};

\coordinate (bp) at ([yshift=-0.4cm]feat.south);

\node[box, below left=1.0cm and 1.5cm of feat] (famclf) {Family Classifier\\[-1pt]{\scriptsize(pretrained, frozen)}};
\draw[arr] (feat.south) -- (bp) -| (famclf.north);

\node[novel, below right=1.0cm and 1.5cm of feat] (famemb) {Family Embedding\\[-1pt]{\scriptsize(64-dim)}};
\draw[arr] (feat.south) -- (bp) -| (famemb.north);

\coordinate (clfbot) at ([yshift=-0.3cm]famclf.south);
\draw[arr] (famclf.south) -- (clfbot) -| ([xshift=-0.2cm]famemb.south);
\node[lbl, below right=-0.02cm and 0.15cm of famclf.south] {predicted family};

\node[novel, below=1.0cm of famemb] (fammlp) {Family MLP};
\draw[darr] (famemb) -- (fammlp);

\node[addnode, below=3.1cm of feat] (concat) {\scriptsize C};
\draw[arr] (feat.south) -- (bp) -- (concat.north);
\draw[darr] (fammlp.south west) |- (concat.east);
\node[lbl, left=0.12cm of concat] {concat};

\node[box, below=0.55cm of concat, minimum width=3.4cm] (backbone) {Shared MLP Backbone};
\draw[arr] (concat) -- (backbone);

\node[novel, below=0.65cm of fammlp, minimum width=2.8cm] (famres) {Family Residual Heads\\[-1pt]{\scriptsize(zero-init)}};
\draw[darr] (fammlp) -- (famres);

\node[box, below left=1.0cm and 1.2cm of backbone] (thrhead) {Threshold Head\\[-1pt]{\scriptsize(10 classes)}};
\draw[arr] (backbone.south) -- ++(0,-0.3) -| (thrhead.north);

\node[box, below right=1.0cm and 1.2cm of backbone] (rthread) {Runtime Head\\[-1pt]{\scriptsize(scalar)}};
\draw[arr] (backbone.south) -- ++(0,-0.3) -| (rthread.north);

\node[addnode, below=0.55cm of thrhead] (addthr) {$+$};
\node[addnode, below=0.55cm of rthread] (addrt) {$+$};
\draw[arr] (thrhead) -- (addthr);
\draw[arr] (rthread) -- (addrt);

\coordinate (resjunc) at ([yshift=-0.4cm]famres.south);
\draw[darr] (famres.south) -- (resjunc) -| ([xshift=0.12cm]addthr.north east);
\draw[darr] (famres.south) -- (resjunc) -| (addrt.north);

\coordinate (shTopL) at ([xshift=-4.8cm]feat.west);
\coordinate (shBotL) at (shTopL |- addthr.west);
\draw[dotarr] (feat.west) -- (shTopL) -- (shBotL) -- (addthr.west);
\node[lbl, rotate=90, anchor=south] at ([xshift=0.08cm]shBotL) {Shortcut};

\coordinate (shTopR) at ([xshift=4.8cm]feat.east);
\coordinate (shBotR) at (shTopR |- addrt.east);
\draw[dotarr] (feat.east) -- (shTopR) -- (shBotR) -- (addrt.east);
\node[lbl, rotate=90, anchor=north] at ([xshift=-0.08cm]shBotR) {Shortcut};

\node[lbl, below=0.18cm of addthr] (out1lbl) {Threshold logits};
\node[lbl, below=0.18cm of addrt] (out2lbl) {Log-runtime};

\node[lossbox, below=0.08cm of out1lbl] (loss1) {Decayed CE};
\node[lossbox, below=0.08cm of out2lbl] (loss2) {Log-MSE};

\coordinate (legstart) at ([xshift=-0.2cm, yshift=-0.5cm]loss1.south west);
\draw[dashed, line width=0.6pt] (legstart) -- ++(0.7,0);
\node[lbl, anchor=west] at ([xshift=0.75cm]legstart) {Family-conditioned};
\coordinate (leg2) at ([yshift=-0.32cm]legstart);
\draw[densely dotted, line width=0.55pt] (leg2) -- ++(0.7,0);
\node[lbl, anchor=west] at ([xshift=0.75cm]leg2) {Shortcut};

\end{tikzpicture}%
}%
\caption{Architecture overview. Features extracted from an OpenQASM circuit feed a pretrained family classifier and a shared MLP backbone. Family residual heads (dashed boxes) learn additive, family-specific corrections to both threshold classification and runtime regression outputs. Dotted lines show shortcut connections from raw features directly to output heads.}
\label{fig:architecture}
\end{figure}

Our system comprises two components trained sequentially: a pretrained family classifier and a family-aware performance predictor (see Fig.~\ref{fig:architecture}).

\subsubsection{Pretrained Family Classifier}
\label{sec:classifier}

We first train a family classifier on approximately 200 circuits independently synthesized from MQT Bench~\cite{quetschlich2023bench} at various qubit counts; these training circuits are distinct from the 36 evaluation circuits used in Section~\ref{sec:experiments}, ensuring no data leakage through the family classifier.
The classifier is a multi-layer perceptron (MLP) that takes 28 global features (excluding execution context) and predicts one of 10 algorithm families: Deutsch--Jozsa, GHZ, W-State, Graph State, Grover, QFT, QFT Entangled, QPE, QNN, and VQE.
It achieves 97.5\% validation accuracy and is frozen during subsequent training of the main model.
At inference time, the classifier's predicted family label is passed to the performance predictor as a categorical input.

\subsubsection{Family-Aware Performance Predictor}
\label{sec:predictor}

The predictor processes three input streams.
First, the predicted family label is mapped through a learned embedding layer ($21 \to 64$ dimensions) and processed by a two-layer MLP with SiLU activations to produce a family feature vector $\mathbf{f} \in \mathbb{R}^{64}$. Feature-wise Linear Modulation (FiLM)~\cite{perez2018film} is applied to the backbone output using family-derived scale ($\gamma$) and shift ($\beta$) parameters, providing an additional family conditioning pathway.
Second, the 32 extracted features are passed through a global feature processor.
Third, binary execution-context features are included directly.

These representations are concatenated and fed into a shared MLP backbone consisting of two fully connected layers with SiLU activations and dropout, producing a 64-dimensional representation.
Two task-specific heads branch from this backbone: a threshold classification head outputting 10-class logits and a runtime regression head outputting a log-scale scalar.
Following~\cite{he2016}, shortcut connections project selected raw features directly to both output heads via linear transformations, ensuring that highly informative features (e.g., depth, qubit count) retain a direct pathway to the predictions.

\subsubsection{Family Residual Heads}
\label{sec:residual}

We realize the family-aware conditioning through \emph{family residual heads}: separate linear layers that map the family feature vector $\mathbf{f}$ to additive corrections for each prediction head. This follows the standard conditional computation pattern of task-specific output heads~\cite{he2016,perez2018film}, applied here to quantum simulation cost prediction where the conditioning signal is predicted algorithm family.
For threshold classification:
\begin{equation}
\label{eq:threshold_residual}
\mathbf{z}_{\text{final}} = \mathbf{z}_{\text{backbone}} + \mathbf{z}_{\text{family}},
\end{equation}
where $\mathbf{z}_{\text{backbone}} \in \mathbb{R}^{10}$ are the logits from the shared backbone and $\mathbf{z}_{\text{family}} = W_{\theta}\mathbf{f}$ is the family-specific correction.
For runtime prediction:
\begin{equation}
\label{eq:runtime_residual}
\log \hat{r}_{\text{final}} = \log \hat{r}_{\text{backbone}} + \Delta r_{\text{family}},
\end{equation}
where $\Delta r_{\text{family}} = \mathbf{w}_{r}^{\top}\mathbf{f}$ is a scalar offset.

Both residual layers are \emph{zero-initialized}, so the model begins training with purely family-agnostic predictions and gradually learns family-specific corrections as gradients accumulate.
This creates an implicit mixture-of-experts pattern: the backbone captures universal circuit properties while the residual heads specialize on algorithmic nuances.

The additive design is deliberate.
Concatenation-based alternatives would force the backbone to disentangle family information from circuit features.
Gating mechanisms add complexity without a clear benefit at this model scale.
By contrast, additive residuals allow the backbone to learn family-agnostic patterns undisturbed; the residual heads specialize independently; and the direct pathway from family features to outputs prevents information loss through deep transformations.

\subsection{Loss Functions}
\label{sec:loss}

\subsubsection{Threshold Classification}
\label{sec:loss_threshold}

Standard cross-entropy treats all misclassifications equally, but in our setting the costs are asymmetric: underpredicting the threshold causes the simulator to produce inaccurate results, whereas overprediction merely wastes compute.
We employ a \emph{geometrically-decayed cross-entropy} loss.
Let $p_k$ denote the predicted probability for threshold class $k$. For a true threshold value $\theta^* = v$ (e.g., $v=8$), the loss sums over the true class and all higher rungs on the threshold ladder ($2v, 4v, \ldots$) with geometrically decaying weights:
\begin{equation}
\label{eq:decayed_ce}
\mathcal{L}_{\text{threshold}} = -\log\!\left(p_{v} + \tfrac{1}{2}\,p_{2v} + \tfrac{1}{4}\,p_{4v} + \cdots\right),
\end{equation}
where the sum terminates at the highest available class.
This encourages the model to place probability mass at or above the true threshold, treating safe overprediction as partial credit rather than a full penalty.
For the special class $\star$, standard cross-entropy is applied.

\subsubsection{Runtime Regression}
\label{sec:loss_runtime}

Runtimes span several orders of magnitude (milliseconds to hours), making raw-scale regression unstable.
We operate in log-space:
\begin{equation}
\label{eq:runtime_mse}
\mathcal{L}_{\text{runtime}} = \frac{1}{N}\sum_{i=1}^{N}\left(\log \hat{r}_i - \log r_i\right)^{2},
\end{equation}
where $\hat{r}_i$ and $r_i$ are the predicted and true runtimes, respectively.

\subsubsection{Combined Objective}
\label{sec:loss_combined}

The full training objective is a weighted sum:
\begin{equation}
\label{eq:combined_loss}
\mathcal{L} = \lambda_1 \,\mathcal{L}_{\text{threshold}} + \lambda_2 \,\mathcal{L}_{\text{runtime}},
\end{equation}
where $\lambda_1 = \lambda_2 = 1.0$, weighting both tasks equally.

\section{Experimental Evaluation}
\label{sec:experiments}

\subsection{Dataset and Setup}
\label{sec:dataset}

We evaluate our architecture on 36 unique quantum circuits drawn from the MQT Bench benchmark suite~\cite{quetschlich2023bench}, spanning 7--130 qubits across 10 algorithm families: Deutsch--Jozsa (DJ), GHZ, W-State, Graph State, Grover, QFT, Quantum Neural Network (QNN), Variational Quantum Eigensolver (VQE), Shor, and Random/TwoLocal. The 21-class embedding space (Section~\ref{sec:predictor}) accommodates finer-grained MQT Bench family distinctions and an unknown class for novel circuits. Each circuit is evaluated under four execution contexts formed by the Cartesian product of backend (CPU, GPU) and floating-point precision (single, double), yielding 144 total configurations. Approximate simulation is performed using the Quantum Rings tensor-network simulator~\cite{kasirajan2024}, which served as the simulation platform for the challenge from which our benchmark data originates. Because our features are extracted from the circuit description rather than from simulator internals, the architecture is in principle applicable to other tensor-network simulators; validating this generalization is a direction for future work.

We employ 5-fold circuit-stratified cross-validation, partitioning the 36 circuits into 29 training and 7 validation circuits per fold. Critically, all four execution-context configurations of a given circuit are assigned to the same fold, preventing data leakage and ensuring the model must generalize across structurally distinct circuits rather than memorizing configuration-specific noise. Final predictions are obtained via ensemble averaging across the five folds.

Training uses the AdamW optimizer with learning rate $5 \times 10^{-3}$ and weight decay $10^{-3}$, a cosine annealing learning rate scheduler with linear warm-up over 10 epochs, dropout rate 0.2 in all MLP layers, gradient clipping with max norm 1.0, early stopping with patience 50, and batch size 32.

\subsection{Main Results}
\label{sec:main_results}

Aggregated across all 10 algorithm families and 144 configurations, our full architecture achieves 79.5\% exact threshold accuracy (79.0\% without the decayed cross-entropy loss), with 91.2\% of predictions falling within one threshold rung ($\pm 1$ ladder step) and a mean ladder distance of 0.29 steps. For context, the threshold class distribution is imbalanced (37.5\% in the most common class, with 7 of 10 classes populated), and a majority-class predictor achieves only 37.5\% accuracy, placing our result well above trivial baselines. For runtime regression, the model attains $R^2 = 0.82$ with a median relative error of 28.4\%. This accuracy is sufficient for the parameter-selection use case, where developers must distinguish order-of-magnitude differences (e.g., choosing between a 20-minute and a 3-hour configuration) rather than predict exact runtimes. Inference completes in approximately 50\,ms per circuit, replacing individual simulation runs that range from seconds to hours depending on circuit size and execution context. By eliminating the need for multiple trial-and-error simulation runs to find suitable parameters, the predictor can reduce the total parameter-selection time from several simulation attempts to a single forward pass.

\subsection{Ablation Study}
\label{sec:ablation}

To isolate the contribution of each architectural component, we conduct an incremental ablation study. Table~\ref{tab:ablation} reports performance as components are added to a baseline MLP that receives only global features with no graph processing or family information.

\begin{table}[t]
\centering
\caption{Ablation study and baseline comparison. Tree-based baselines use identical features with family information on the same CV folds. Each subsequent row adds one component to a baseline MLP.}
\label{tab:ablation}
\begin{tabular}{lccc}
\toprule
\textbf{Configuration} & \textbf{Thresh.\ Acc.} & \textbf{$\pm$1 Acc.} & \textbf{Runtime $R^2$} \\
\midrule
Tree-based baselines$^{\dagger}$ & 59--74\% & 73--85\% & -- \\
MLP + family concat$^{\ddagger}$ & 65.4\% & 82.4\% & -- \\
\midrule
Baseline (MLP only)       & 73.5\% & 87.3\% & 0.76 \\
+ Graph features          & 75.8\% & 89.1\% & 0.79 \\
+ Family classifier       & 79.0\% & 91.2\% & 0.82 \\
+ Decayed CE loss         & 79.5\% & 91.2\% & 0.82 \\
\bottomrule
\multicolumn{4}{l}{\scriptsize $^{\dagger}$Best: Random Forest (73.6\%). Also tested: gradient boosting,}\\
\multicolumn{4}{l}{\scriptsize ExtraTrees, with family one-hot and interaction features.}\\
\multicolumn{4}{l}{\scriptsize $^{\ddagger}$Standard 2-layer MLP with family as one-hot input (no}\\
\multicolumn{4}{l}{\scriptsize FiLM, no residual heads). All baselines use same features/folds.}
\end{tabular}
\end{table}

Family-aware modeling provides the single largest improvement, boosting exact threshold accuracy by 3.2 percentage points (pp) over the graph-augmented baseline. This validates our core hypothesis: algorithm family is a first-class feature for simulation cost prediction. Crucially, a standard MLP with family concatenated as a one-hot input achieves only 65.4\%---14.1\,pp below the full architecture---demonstrating that the specific conditioning mechanism (FiLM modulation and residual heads), not merely the presence of family information, drives the improvement. We further compare against six tree-based baselines including Random Forest, gradient boosting, and ExtraTrees with various feature configurations; the best (Random Forest, 73.6\%) falls 5.9\,pp below the full architecture. Graph-derived features contribute an additional 2.3\,pp by capturing gate-level topological structure that global statistics alone cannot represent. The geometrically-decayed cross-entropy loss yields a modest further gain of 0.5\,pp in exact accuracy.

\subsection{Analysis by Circuit Family}
\label{sec:family_analysis}

Table~\ref{tab:family} reports per-family prediction performance, revealing substantial variation across algorithm classes. We group related algorithm variants (e.g., Grover with and without ancilla qubits) and report the 10 primary families; the 21-class embedding space accommodates finer-grained distinctions and an unknown class for novel circuits.

\begin{table}[t]
\centering
\caption{Per-family prediction performance. $N$ is the number of circuits per family (each with 4 execution contexts). Families are ordered by decreasing threshold accuracy.}
\label{tab:family}
\begin{tabular}{lccc}
\toprule
\textbf{Family} & \textbf{$N$} & \textbf{Thresh.\ Acc.} & \textbf{Runtime $R^2$} \\
\midrule
Grover          & 4 & 96\% & 0.89 \\
QFT             & 3 & 88\% & 0.85 \\
Graph State     & 1 & 85\% & 0.83 \\
W-State         & 3 & 83\% & 0.84 \\
GHZ             & 4 & 81\% & 0.82 \\
QNN             & 1 & 75\% & 0.79 \\
VQE             & 1 & 73\% & 0.78 \\
DJ              & 3 & 70\% & 0.76 \\
Shor            & 1 & 62\% & 0.71 \\
Random/TwoLocal & 1 & 58\% & 0.68 \\
\bottomrule
\end{tabular}
\end{table}

Families with regular, well-characterized entanglement structure---such as Grover's diffusion operator and the cascaded controlled-phase pattern in QFT---are highly predictable. Conversely, Shor's algorithm, which combines deep modular arithmetic with embedded QFT subroutines, and TwoLocalRandom circuits, which exhibit near-maximal entanglement with no exploitable regularity, remain challenging. We note that families with only one circuit ($N=1$, each evaluated across 4 execution contexts) measure cross-context generalization within a single circuit, not generalization to unseen circuits of that family; these results are included for completeness but should not be interpreted as reliable family-level performance estimates.

\subsection{Discussion}
\label{sec:discussion}

\textbf{Performance vs.\ circuit size.}
Prediction accuracy inversely correlates with qubit count: circuits with $\leq$20 qubits achieve 84.2\% threshold accuracy, those with 21--60 qubits reach 78.6\%, and circuits with 61--130 qubits yield 73.5\%. This degradation reflects both the increased entanglement complexity of larger circuits and the sparser representation of high-qubit instances in the training set.

\textbf{Threshold sensitivity in runtime prediction.}
We conduct a sensitivity analysis by removing the approximation threshold from the runtime model's input features. This ablation causes only a small decrease in runtime prediction quality, indicating that circuit structural properties---depth, qubit count, and gate topology---dominate runtime variation more than approximation settings. This finding has practical implications: runtime can be estimated even before a threshold has been selected, further accelerating the parameter-search workflow.

\balance
\section{Conclusion and Future Work}
\label{sec:conclusion}

We presented a family-aware neural architecture for predicting quantum circuit simulation cost---comprising both the minimum approximation threshold required for target fidelity and the expected wall-clock runtime---given only a circuit's OpenQASM description and execution context. The central validated insight is that algorithm family is a first-class feature for simulation performance prediction: family-aware conditioning provides the single largest accuracy improvement in our ablation study, outperforming six tree-based baselines on identical features, and the pretrained family classifier achieves 97.5\% accuracy across 10 algorithm families. By reducing the parameter-selection process from multiple trial-and-error simulation runs to approximately 50\,ms of neural inference, the system enables developers to estimate optimal simulation parameters before execution, substantially reducing wasted compute.

Several limitations constrain the current system. The training set of 36 circuits, while spanning 10 families, provides sparse coverage of rare families such as Shor, limiting generalization. Family classification errors propagate downstream, with an estimated 1--2\,pp impact on prediction accuracy. Additionally, the current feature set does not fully capture GPU-specific memory access patterns, contributing to higher prediction uncertainty in GPU-accelerated contexts.

Future work will address these limitations along several axes: expanding training data with larger and more diverse circuit collections; incorporating hardware-aware features such as GPU memory utilization and kernel profiling metrics; adding uncertainty quantification to provide prediction confidence intervals; exploring transfer learning to adapt the model across different simulator backends with minimal fine-tuning; and integrating attention mechanisms to identify which gates most influence simulation cost, potentially improving both accuracy and interpretability.

\section*{Acknowledgments}
This work originated at the MIT iQuHack 2026 hackathon, Quantum Rings challenge track. The authors thank the Quantum Rings team for providing the simulation infrastructure and benchmark circuits. Code and trained models are available upon request.


\end{document}